# Designing flexible hard magnetic materials for zero-magnetic-field operation of the anomalous Nernst effect

Sang J. Park[1,*], Rajkumar Modak[1,#], Ravi Gautam[1], Abdulkareem Alasli[2], Takamasa Hirai[1], Fuyuki Ando[1], Hosei Nagano[2], Hossein Sepehri-Amin[1] and Ken-ichi Uchida[1,3,*]

[1] National Institute for Materials Science, Tsukuba 305-0047, Japan

[2] Department of Mechanical Systems Engineering, Nagoya University, Nagoya 464-8601, Japan

[3] Department of Advanced Materials Science, Graduate School of Frontier Sciences, The University of Tokyo, Kashiwa 277-8561, Japan

[#]Present affiliation: Department of Advanced Materials Science, Graduate School of Frontier Sciences, The University of Tokyo, Kashiwa 277-8561, Japan

*Correspondence to: PARK.SangJun@nims.go.jp (S.J.P.); UCHIDA.Kenichi@nims.go.jp (K.U.)

## Abstract

The global shift towards a carbon-neutral society has accelerated the demand for green energy, driving research into efficient technologies for harvesting energy from low-grade waste heat. Recently, transverse thermoelectrics based on the anomalous Nernst effect (ANE) has gained attention due to their simple device structure, scalability, and manufacturing-friendly nature. While topological single crystals and epitaxial films have been focused for enhancing the ANE-driven thermoelectric performance, further improvements in material design are necessary for practical applications. Here, we report an easy-to-implement strategy for designing mechanically flexible and magnetically hard transverse thermoelectric materials by creating amorphous-crystalline heterogeneous composites. We fabricated and optimized the heterogeneous composites through controlled heat treatment, achieving significant enhancements in the coercivity and anomalous Nernst coefficient, while maintaining flexibility. Additionally, using the developed material, we constructed a single-material-based coiled device and demonstrated the zero-field operation of the ANE-based energy harvesting from curved heat sources. These results validate the feasibility of using the ANE-based flexible materials for energy harvesting applications.

## 1. Introduction

Given that a significant portion of energy is dissipated into the environment as heat during the energy conversion process (e.g., approximately 66% for fossil fuel combustion), recovering such waste heat is of great importance for achieving a sustainable future[1,2]. One potential method to utilize the low-grade waste heat for generating usable electricity is thermoelectric generation, which has been extensively studied due to its various advantages, such as soundless operation, low maintenance, and the absence of moving parts and hazardous working fluids[3–7]. Over the past century, thermoelectric generation has been developed based on longitudinal geometry using the Seebeck effect, wherein a charge current is generated parallel to the input heat flux[3–7]. Typical longitudinal thermoelectric devices contain multiple pairs of n- and p-type conductor legs connected electrically in series and thermally in parallel, allowing the cumulative increase in output electrical voltage. However, such multiple $\pi$-shaped junctions render the structure of thermoelectric devices intricate, thus limiting scalability, manufacturability, flexibility, and tunability of devices.

Recently, energy harvesting based on transverse thermoelectric effects has attracted attention to overcome the limitations of longitudinal devices[8–12]. In transverse thermoelectric conversion geometry, a charge current is generated in the direction perpendicular to the heat flux, which enables the construction of much-simplified device structures. Specifically, transverse thermoelectric devices can be constructed using a single material without complex pairs and junctions (Fig. 1). Such a junction-free design minimizes performance degradation at the device level due to electrical and thermal contact resistances and mass diffusion between thermoelectric legs and electrodes at the hot side, which are the bottlenecks for practical applications[8,13]. As a result, the efficiency of the transverse thermoelectric device can be closer aligned with the intrinsic properties of materials. The output of the transverse thermoelectric effects can also be enhanced by designing device dimensions; the voltage and power can be increased by increasing the length and area of the material, respectively, in the direction perpendicular to the heat flux, which makes transverse thermoelectrics attractive for applications despite its lower thermoelectric conversion efficiency compared to longitudinal counterparts[8–11,14–16].

The Nernst effect is a well-known transverse thermoelectric effect, where a charge current is generated orthogonally to both an applied heat flux and magnetic field in a nonmagnetic conductor, specifically termed the ordinary Nernst effect (ONE). In a conductor with spontaneous magnetization (e.g., ferromagnet), an additional transverse thermoelectric effect can occur depending on the intrinsic magnetization of the conductor, a phenomenon called the anomalous Nernst effect (ANE)[11]. ANE-based transverse thermoelectric devices can operate without an external magnetic field when the material has finite coercivity and remanent magnetization. ANE allows for a streamlined device structure compared to ONE that always requires an external magnetic field and/or additional hard magnets[17]. To date, many researchers have focused on developing new topological materials with large ANE, exhibiting non-trivial band topologies[8,9,11]. Examples include magnetic Heusler alloy $Co_2MnGa$[18–20], Kagome ferromagnets $Co_2Sn_2S_2$[21] and $UCo_{0.8}Ru_{0.2}Al$[22], binary Fe-based alloys[23,24], antiferromagnetic $YbMnBi_2$[25], which demonstrate large ANE by the Weyl nodes, lines, and nodal webs near the Fermi level, yielding a large Berry curvature.

Although significant improvements in ANE have been made through an improved understanding of band topology, further progress is needed for practical applications of ANE-based transverse thermoelectric devices due to several challenges. First, most of the aforementioned research has focused on single crystals or epitaxial thin films, which are useful for investigating the unique physics of transverse transport properties arising from band topology or magnons[26]. However, these materials are not ideal for applications due to their high material cost, mechanical instability, and limited scalability[27–30], which paradoxically make practical applications more challenging than Seebeck-based longitudinal thermoelectric devices. Thin films are particularly unsuitable for extracting substantial power because their small volume limits the amount of energy that can be harvested. Additionally, the good transport properties related to ANE of these materials require high-quality, clean samples due to the sensitivity of ANE to the Fermi level[20,31,32]. Second, most of the ANE materials above are magnetically soft and possess small coercivity (typically on the order of a few tens of Oe). Such low coercivity poses a challenge in guaranteeing device performance in the absence of an external magnetic field, undermining one of the key advantages of using ANE over ONE. Finally, the rigidity of these systems poses another limitation; one potential advantage of ANE-based thermoelectric devices is their ability to operate on non-flat surfaces, which could enable energy harvesting from heat sources with curved or rough surfaces[33–35]. Considering available heat sources in a society such as heat pipes, flexible devices are more useful for energy harvesting[14,36–38].

For their practical applications, it is necessary to develop ANE materials or devices that combine large ANE-based thermoelectric conversion performance, high coercivity, and flexibility, without using single-crystalline materials. While some of these aspects have been researched separately, no studies have successfully addressed all of them together in a comprehensive manner. Here, we propose a strategy for designing ANE materials that meet these criteria. We experimentally validate this material design using an amorphous-crystalline heterogeneous composite as a platform. We optimized and fabricated a heterogeneous composite material consisting of amorphous and crystalline phases (Fig. 1), which impart flexibility and coercivity to the systems, respectively. Our samples demonstrated a large coercivity $H_C$ of 750-950 Oe and anomalous Nernst coefficient of $2.73 \pm 0.41$ µV/K while maintaining mechanical flexibility. Furthermore, we experimentally validated the potential applicability of these ANE materials for energy harvesting from curved heat sources by constructing a coiled device (Fig. 1).

To provide a comprehensive understanding of our research, this paper is organized as follows. We first outline the material design strategy for creating flexible magnetic materials with large coercivity and enhanced ANE, along with its experimental demonstration using amorphous-crystalline heterogeneous composites (Section 2.1). Next, we present the characterization of these materials, focusing on the relationship between the crystal structure and enlarged coercivity (Section 2.2), as well as their transverse thermoelectric performance (Section 2.3). We then explore the application of these materials in a coiled device structure designed for energy harvesting from curved heat sources (Section 2.4). A comparative analysis is provided to evaluate the performance of these materials against other flexible and rigid systems, highlighting the advantages of our approach (Section 2.5). Finally, we summarize the key findings and discuss their implications for future research and development in energy harvesting technologies (Section 3).

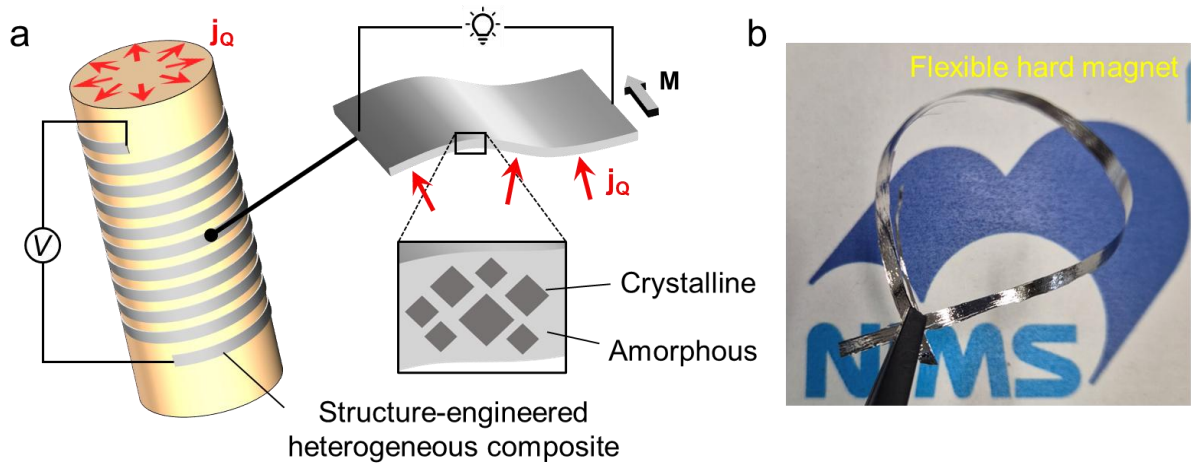

**Fig. 1 | Flexible hard magnets for zero-field operation of the anomalous Nernst effect.** (a) Schematic illustration of a transverse thermoelectric device operated in zero magnetic field, utilizing an amorphous-crystalline heterogeneous composite material on a curved heat source. The coiled device structure enables cumulative increases in electrical voltage along the device length in a response to the radial heat flux $\mathbf{j}_Q$. $\mathbf{M}$ denotes the spontaneous magnetization of the material. (b) Photograph of the fabricated $Fe_{63}Pt_{15}B_{10}Si_{12}$ flexible hard magnet annealed at 873 K for 3 min.

## 2. Results and discussion

### 2.1 Designing flexible ferromagnetic materials with large coercivity by controlling crystal structures

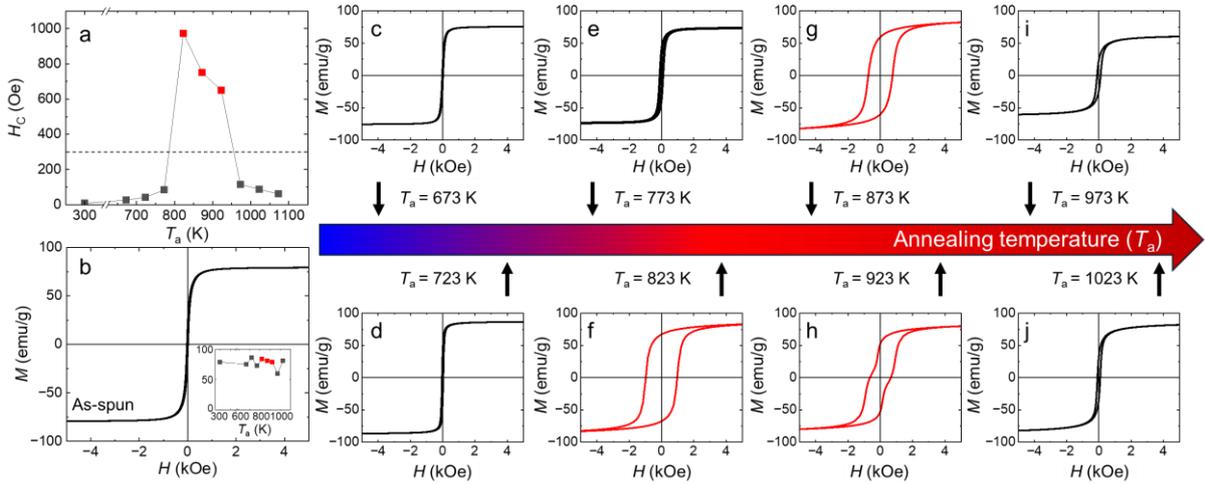

**Fig. 2 | Designing flexible ferromagnetic materials with large coercivity by controlling annealing temperatures ($T_a$).** (a) $T_a$ dependence of the coercivity of the $Fe_{63}Pt_{15}B_{10}Si_{12}$ samples at room temperature. The horizontal dashed line in (a) indicates the lower limit of coercivity used in credit cards[39]. Magnetic properties of (b) as-spun without annealing, (c) 673 K, (d) 723 K, (e) 773 K, (f) 823 K, (g) 873 K, (h) 923 K, (i) 973 K, and (j) 1023 K. The inset in (b) indicates the saturation magnetization of the samples as a function of $T_a$. The data for $T_a = 300$ K in the inset correspond to the as-spun sample. The magnetic properties were measured at room temperature. The red symbols in (a) and (b), and red lines in (f)-(h) indicate the properties of the samples exhibiting a significant enhancement of the coercivity.

First, we fabricated an Fe-based amorphous metal using the melt spinning process, with a nominal composition of $Fe_{63}Pt_{15}B_{10}Si_{12}$ (at.%) and a thickness of 29.5±3.4 μm (Fig. 1b, see Methods for details). The addition of B and Si to Fe in the rapidly solidified ribbons is well known for forming an amorphous phase with mechanical flexibility while possessing soft magnetic properties[40]. To enable zero-field operation of ANE, 15 at.% of Pt was added; this addition was intended to increase $H_C$ by forming Fe-Pt-based crystalline phases through annealing[41–43], such as L1$_0$-FePt[44–46], γ-FePt[43,47] or Fe$_3$Pt[39], all of which are well known to possess large $H_C$ due to the large magnetocrystalline anisotropy derived from spin-orbit interaction. Additionally, an improvement in transverse transport properties was also anticipated due to the large spin-orbit coupling of the heavy element Pt in a microscopic manner[48–50] and recently proposed side-jump-like macroscopic transverse scattering in heterogeneous composites[30]. The annealing process was conducted at various annealing temperatures $T_a$ to systematically control the structure and crystalline phases of the materials. The samples were water-quenched after annealing at each $T_a$ for 3 min, which was a sufficiently short time for crystal growth. This short annealing time was designed to form an amorphous-crystalline heterogeneous composite structure, providing both mechanical flexibility and large coercivity within the materials.

The change in $H_C$ was measured using a vibrating sample magnetometer (VSM) at room temperature. The magnetic field was applied in the in-plane direction aligning with the measurement configuration of transverse thermoelectric generation, as shown later (Sections 2.4 and 2.5). Figure 2b shows the magnetic hysteresis of the as-spun sample without annealing, exhibiting soft ferromagnetic properties with small $H_C < 10$ Oe. Upon annealing at various values of $T_a$, a drastic increase in $H_C$ occurred, while the saturation magnetization $M_s$ remained at similar values (Figs. 2a and 2b inset). Notably, the samples annealed at $T_a = 823$, 873, and 923 K showed significantly large $H_C$, in agreement with previous observations with similar compositions[41–43]. The $H_C$ values exhibited a sharp increase between $T_a = 773$ K and 823 K, consistent with the crystallization temperature (Fig. 3a), indicating a direct relation to the formation of crystalline phases. The $H_C$ value reached a maximum in the $T_a = 823$ K sample (i.e., $H_C = 973$ Oe) and slightly decreased in the samples annealed at $T_a = 873$ K (i.e., $H_C = 750$ Oe) and 923 K (i.e., $H_C = 650$ Oe). Above $T_a = 923$ K, $H_C$ drastically decreased. Notably, the $T_a$ = 823 K and 873 K samples showed large remanence ($M_r$) with good squareness (Figs. 2f and 2g). It is noteworthy that, in addition to large $H_C$, a high remanence ratio (=$M_r/M_s$) is important to ensure the performance of ANE in zero-field operation. The $T_a$ = 823 K and 873 K samples possess high $M_r/M_s$ ratios of 78 % and 73 %, respectively, both of which are much larger than that in previously reported Fe₃Pt thin films ($M_r/M_s$ ~ 17 % with $H_C$ ~ 1300 Oe[ref.[39]]) grown on rigid MgO substrates.

On the other hand, it has been reported that the annealing process reduces the ductility of amorphous alloys, thereby decreasing their flexibility[40,41]. The change in mechanical properties can be attributed to a reduced portion of the ductile amorphous phase, as the brittle crystalline portion increases with $T_a$. To evaluate flexibility, we attached the annealed samples to a curved surface with a radius of 5 mm (Fig. S1 in Supplementary Information). Notably, the materials studied here exhibited high flexibility, suggesting the potential applicability of these materials to energy harvesting on curved heat sources.

To further clarify the origin of enhanced $H_C$, we prepared an additional sample set annealed at the same temperature conditions, from 673 K to 1023 K with 50 K increments, but for a longer holding time of 30 min with natural cooling inside the furnace. The additional sample set was anticipated to possess a larger crystalline size compared to the main sample set. In the 30-min annealed samples, we consistently observed large $H_C$ values of 670-1170 Oe (Fig. S2 in Supplementary Information). The maximum $H_C$ of 1170 Oe in the $T_a$= 973 K sample is comparable to the Fe₃Pt thin film on a rigid MgO substrate[39]. Overall, the 30-min annealed sample set exhibited superior $H_C$ compared to the 3-min annealed sample set, supporting that the increased $H_C$ is attributed to the intricate change in microstructural features. This behavior is consistent with the previous observation in Fe-Pt-Si systems[51], in which the maximum $H_C$ occurred with a longer annealing time of 3 h.

Despite the large $H_C$ in the 30-min annealed samples, this sample set was not suitable for applications on curved heat sources because of their brittleness. We confirmed that the 30-min annealed samples lost flexibility, possibly due to the lack of an amorphous portion that gives flexible functionality to the system. Therefore, from now on, we focus on the sample set annealed for 3 min that exhibited large $H_C$ (650-973 Oe) while maintaining flexibility. It is noted that the observed $H_C$ in the 3-min annealed samples is still sufficiently large compared to the reported ANE materials, including rigid single crystals ($H_C$ < 60 Oe), thin films on rigid substrates ($H_C$ < 60 Oe for typical cases[18–20,23,52–54] and $H_C$ ~ 1300 Oe for $Fe_3Pt$ epitaxial film[39]), non-single crystal films on flexible substrates ($H_C$ < 80 Oe for polycrystalline films[34,35,55,56] and $H_C$ ~ 320 Oe for Sm-Co-based amorphous alloy[57]), and flexible bulk amorphous metals ($H_C$ < 10 Oe)[40,58]. Although the bulk permanent magnet $SmCo_5$ exhibits a much larger coercivity of $H_C$ ~ 25 kOe[ref.59], it is rigid and not applicable to curved heat sources. A detailed comparison is shown later, together with ANE performance (Section 2.5).

## 2.2 Crystal structure analysis

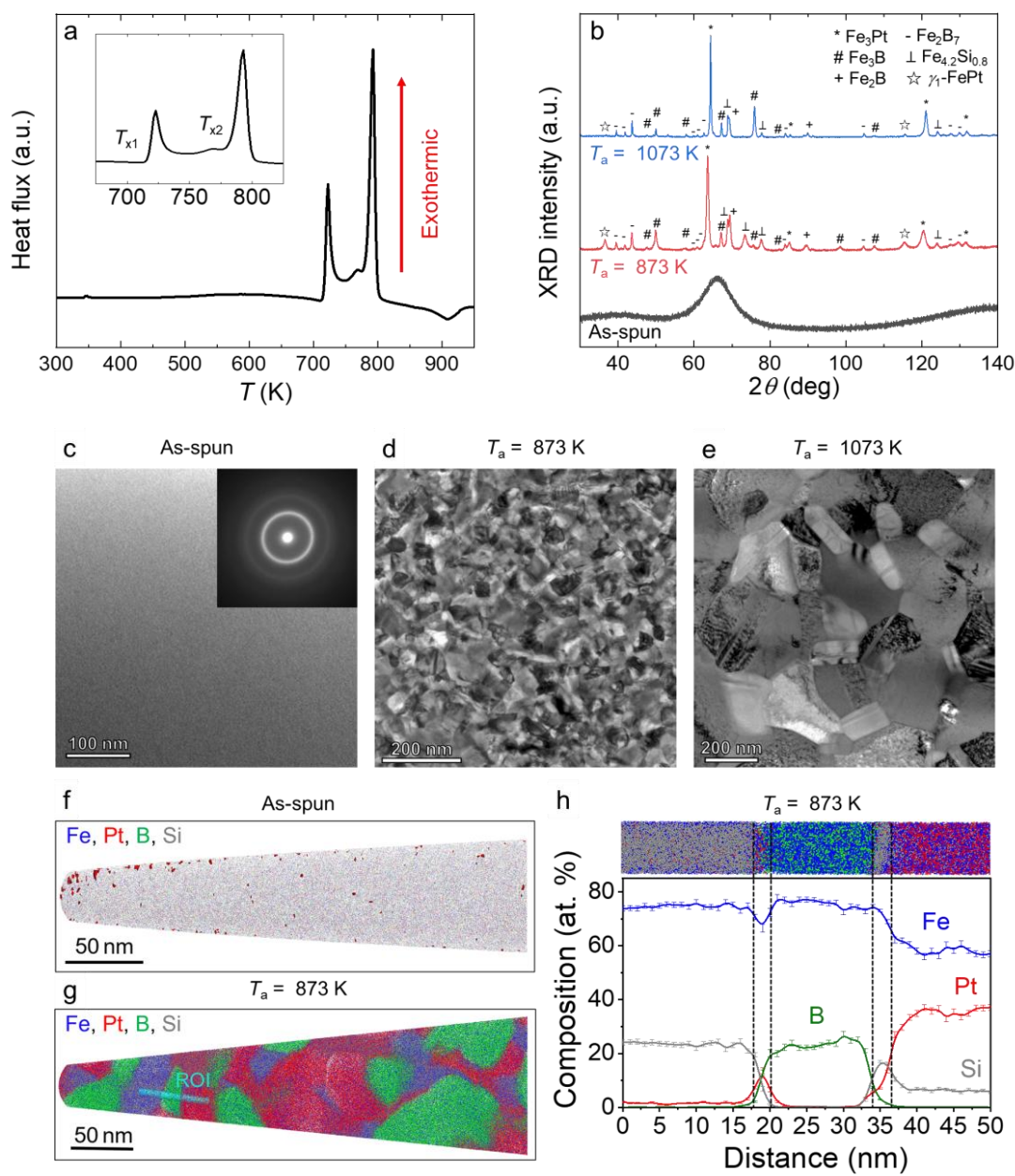

**Fig. 3 | Investigation of crystallinity change in the heterogeneous samples.** (a) Exothermic reactions during heating measured by differential scanning calorimetry. Two distinct peaks were observed at $T_{x1}$ = 722 K and $T_{x2}$ = 792 K. (b) X-ray diffraction patterns of the as-spun and annealed samples at $T_a$ = 873 K and 1073 K. (c)-(e) Low-magnification transmission electron microscopy images of the as-spun sample (c) and the annealed samples at $T_a$ = 873 K (d) and 1073 K (e). (f), (g) Atom probe tomography images of the as-spun sample (f) and the annealed sample at $T_a$ = 873 K (g). (h) Elemental map and corresponding composition line profile of the constituent elements obtained from a region of interest (ROI) shown in (g).

We now focus on the crystal structure of the $H_C$-enhanced samples to further clarify the origin of the enhancement. Initially, the dynamic change in crystallinity was investigated using differential scanning calorimetry (DSC). The DSC data in Fig. 3a show two distinct exothermic reactions at $T_{x1} = 722$ K and $T_{x2} = 792$ K during heating from 300 K to 990 K at a ramp rate of 10 K/min, a characteristic of the crystallization[41,60,61]. Considering the previously observed transition temperature in $H_C$ (773 K < $T$ < 823 K, Fig. 2c), the occurrence of large $H_C$ is more likely associated with $T_{x2}$, which corresponds to the formation of Fe-based metalloid phases[62]. Next, structural analyses were performed using X-ray diffraction (XRD). Three representative samples were selected for this investigation: the as-spun, $T_a = 873$ K, and 1073 K samples, which have $H_C$ of < 10 Oe, 750 Oe, and 88 Oe, respectively. Although the $T_a = 823$ K sample exhibited the maximum $H_C$, the $T_a = 873$ K sample was chosen for structural investigation considering its large ANE performance, shown in the following section (Fig. 4). The other two samples serve as control, representing non-annealed (as-spun) and fully crystallized ($T_a = 1073$ K) conditions. Figure 3b shows that the $T_a = 873$ K and 1073 K samples possess similar crystal structures, including $Fe_3Pt$ and $\gamma_1$-FePt, while the as-spun sample exhibits an amorphous structure with a broad XRD peak. The only distinguishable difference between the $T_a = 873$ K and 1073 K samples from the XRD data is that the former exhibits a superior $\gamma_1$-FePt, known to have large $H_C$[ref.[43,47]].

To further clarify the correlation between $H_C$ and crystal structure, transmission electron microscopy (TEM) was utilized on the same sample set. Figures 3c-e show the formation of crystalline grains in the amorphous structure upon annealing, while the crystalline size gradually increases with increasing $T_a$. The as-spun sample shows a halo ring diffraction pattern due to the amorphous nature, as expected (Fig. 3c inset). Notably, the $T_a = 873$ K sample exhibits sub-100-nm scale grains, while the $T_a$ = 1073 K sample has a larger grain size. The sub-100-nm scale grains are known to be beneficial for large $H_C$, following the general trend observed in soft magnetic amorphous and polycrystalline alloys[63]. In addition, the atom probe tomography (APT) was employed to investigate the nanostructures of the $T_a$ = 873 K sample. Figures 3f-h clearly reveal the nanocrystals at the grain boundaries, which may provide sources for domain wall pinning[42]. Specifically, the composition line profile, obtained from the region of interest (ROI) in Fig. 3g, exhibits the presence of primary nano-scale Fe-Pt- and Fe-Si-based alloy phases at the grain boundaries (highlighted with horizontal dashed lines in Fig. 3h). These phases correspond to $Fe_3Pt$, $\gamma_1$-FePt, and $Fe_{4.2}Si_{0.8}$, as indicated by the XRD data.

Taken together with the measured XRD, TEM, and APT data, we conclude that the $T_a$ = 873 K sample has the favorable crystal structures and grains to produce large $H_C$, thus confirming the potential applicability of structure-engineered amorphous-crystalline heterogeneous composites for ANE-based energy harvesting through spontaneous magnetization at zero magnetic field. In the following section, we discuss the ANE performance to further assess the feasibility of these materials for transverse thermoelectric conversion.

## 2.3 Enhancing transverse thermoelectric conversion in heterogeneous composites

We now present the transverse thermoelectric conversion performance of the samples. We employed lock-in thermography (LIT) to characterize the anomalous Ettingshausen effect (AEE), a transverse charge-to-heat current conversion that is the Onsager reciprocal of ANE. LIT captures the first harmonic response of the temperature change induced by a periodic charge current in terms of the lock-in amplitude $A$ and phase $\phi$. By performing LIT measurements with applying a steady magnetic field and square-wave-modulated AC charge current to a magnetic material, we can detect the AEE-induced temperature change (Fig. 4). The heat current density driven by AEE can be expressed as[59,64–66]

$$\mathbf{j}_{\mathrm{Q,AEE}} = \Pi_{\mathrm{AEE}}(\mathbf{j}_{\mathrm{C}} \times \mathbf{m}), \tag{1}$$

where $\Pi_{\mathrm{AEE}}$, $\mathbf{j}_{\mathrm{C}}$, and $\mathbf{m}$ indicate the anomalous Ettingshausen coefficient, charge current density with the square-wave intensity $j_{\mathrm{C}}$ and unit vector of the magnetization, respectively. Specifically, the infrared (IR) camera measures the IR intensity thermally emitted from the sample surface, which is coated by electrically insulating black ink with a high emissivity over 0.94 (Fig. 4a). Through signal processing incorporating the reference signal, $A$ and $\phi$ can be determined for each pixel in the field-of-view. The temperature modulation in response to a periodic charge current includes the AEE ($\propto j_{\mathrm{C}}$), Peltier effect ($\propto j_{\mathrm{C}}$) and Joule heating ($\propto j_{\mathrm{C}}^2$) contributions. The Joule heating contribution can be extracted by applying a zero-offset square-wave-modulated current (Fig. 4a), generating a constant background signal ($\propto j_{\mathrm{C}}^2$) without any periodic temperature changes. Subsequently, AEE and the magnetic-field-independent Peltier effect can be separated by symmetrizing the signals for the magnetic field[59,64–68]. The AEE-induced temperature change is directly represented by the field-odd components of $A$ and $\phi$, termed $A_{\mathrm{odd}}$ and $\phi_{\mathrm{odd}}$, respectively.

The $\Pi_{\text{AEE}}$ values estimated from the LIT measurements allow us to estimate the anomalous Nernst coefficient $S_{yx}$ quantitatively through the Onsager reciprocity. The anomalous Ettingshausen coefficient measured by LIT is expressed as

$$\Pi_{\text{AEE}} = \frac{\pi}{4} \frac{\kappa \Delta T_{\text{AEE}}}{j_c L},$$ (2)

where $\Delta T_{\text{AEE}}$ is the temperature difference along the thickness ($L$) direction by AEE (i.e., $\Delta T_{\text{AEE}} = 2A_{\text{odd}}$) and $\kappa$ is the thermal conductivity of the sample[59,64–66]. The reciprocal relation gives the anomalous Nernst coefficient as

$$S_{\text{ANE}} = \Pi_{\text{AEE}}/T,$$ (3)

where $T$ is the absolute temperature.

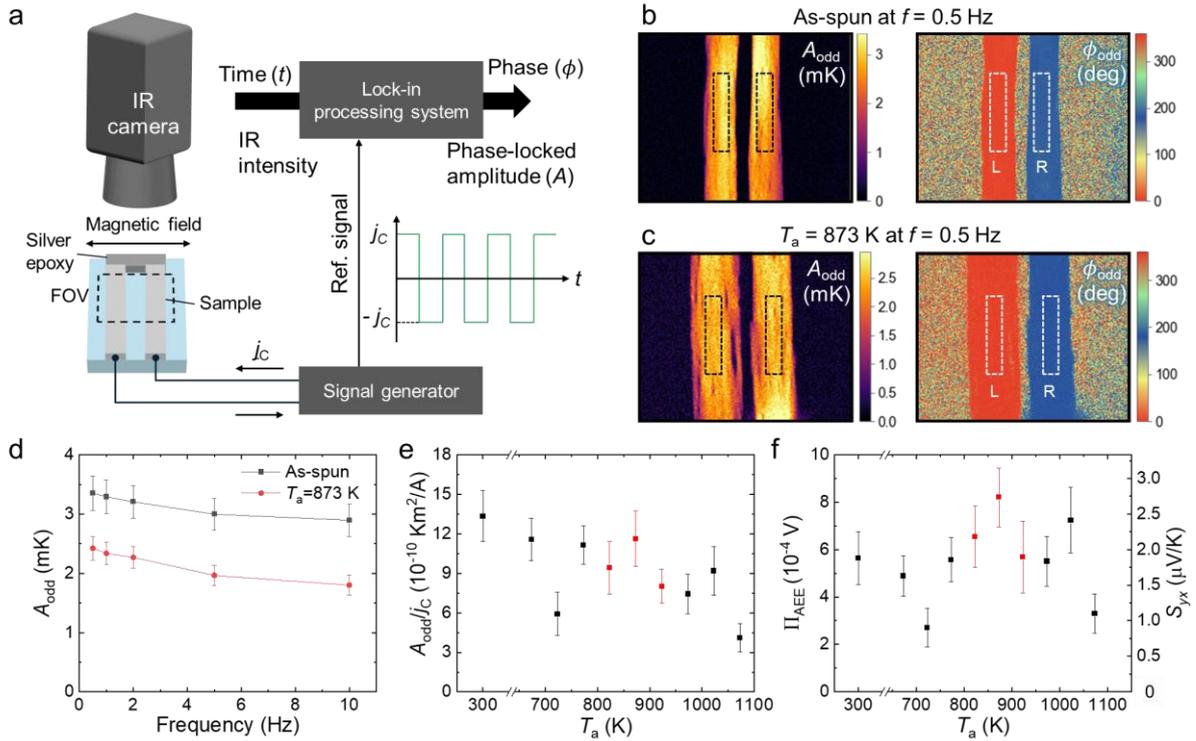

**Fig. 4 | Evaluation of transverse thermoelectric performance of flexible bulk magnetic materials using lock-in thermography (LIT).** (a) Schematic illustration of the LIT technique. (b), (c) Lock-in amplitude ($A_{\text{odd}}$) and phase ($\phi_{\text{odd}}$) images for the as-spun sample (b) and the

$T_a$=873 K sample (c) at a frequency $f$ of 0.5 Hz. (d) $f$ dependence of $A_{odd}$ for the as-spun and $T_a$=873 K samples. (e) $A_{odd}/j_C$ at $T = 300$ K as a function of annealing temperature $T_a$. (f) Anomalous Ettingshausen coefficient $\Pi_{AEE}$ and corresponding anomalous Nernst coefficient $S_{yx}$ (=$\Pi_{AEE}/T$) at $T = 300$ K as a function of $T_a$. The data for $T_a = 300$ K in (e) and (f) correspond to the as-spun sample. The error bars represent the standard deviation of the measurements.

First, we present the $A_{odd}$ and $\phi_{odd}$ images of two representative samples: as-spun and $T_a = 873$ K (Fig. 4b and 4c). The measurements were conducted under an in-plane magnetic field of 6.5 kOe, which is sufficiently large for fully magnetizing both samples, enabling quantitative evaluation of the ANE properties. Two strips of the samples were placed in parallel on a glass substrate and electrically connected in series, yielding opposite current directions between them (Fig. 4a). Considering the symmetry condition of AEE in Eq. (1), temperature gradients of the strips should be opposite, which can be captured by the difference in $\phi_{odd}$. The $\phi_{odd}$ images in Fig. 4b and 4c showed the difference in $\phi_{odd}$ between L and R areas to be ~ 180°, indicating that $A_{odd}$ arises from AEE clearly[59,64–66]. The $A_{odd}$ signal in Fig. 4d was obtained by averaging the values at the center of the samples with flat surfaces (boxed in Fig. 4b and 4c), where the defocusing-induced error is minimized. To quantitatively estimate $\Pi_{AEE}$, we measured the lock-in frequency ($f$) dependence of $A_{odd}$ over a range from 0.5 Hz to 10 Hz. Figure 4d shows that $A_{odd}$ exhibits a small dependence on $f$ below 2 Hz with changes of <7%, indicating a near-steady state. Thus, for $\Pi_{AEE}$ estimation, we used the values at $f = 0.5$ Hz.

Figure 4e shows the $T_a$ dependence of $A_{odd}/j_C$, exhibiting a decreasing trend. The $\Pi_{AEE}$ and $S_{yx}$ values were calculated using Eqs. (2) and (3) with appropriate parameter values, including the thermal conductivity ($\kappa$) and $T = 300$ K. The $\kappa$ value was estimated by multiplying the mass density $\rho$ (= 9.75 g/cm$^3$), specific heat $c_p$ (= 0.391 J/(g·K)), and thermal diffusivity $D$, measured by Archimedes' principle, DSC, and LIT, respectively (see Methods and Fig. S3 in Supplementary Information for details). Figure 4f shows the estimated $\Pi_{\text{AEE}}$ and $S_{yx}$ values in the samples as a function of $T_{\text{a}}$, exhibiting non-monotonic behaviors; $S_{yx}$ peaks at $T_{\text{a}} = 873$ K and decreases at higher $T_{\text{a}}$. Specifically, the $S_{yx}$ value in the $T_{\text{a}} = 873$ K sample reach to $2.73 \pm 0.41$ μV/K, which is approximately 45% and 148% higher than those in the as-spun ($S_{yx} = 1.88 \pm 0.37$ μV/K) and $T_{\text{a}} = 1073$ K ($S_{yx} = 1.10 \pm 0.28$ μV/K) samples, respectively. Such an enhanced behavior with a mild heat treatment condition is consistent with the recent experimental observation in soft magnetic systems ref.[30] and ref.[40]. The enhancement in our system is more relevant to the phenomenon proposed in ref.[30], rather than that in ref.[40], later of which explains the enhancement by precipitated Cu nanocrystals, considering the absence of Cu in our samples. According to the theoretical model on transverse transport properties in heterostructures recently proposed by Park et al.[30], the transverse scattering can occur if both constituent physical mixtures (e.g., amorphous and crystalline) possess significantly different transverse transport properties due to the occurrence of a wiggy path of electrons. Similarly, our samples can be regarded as a physical mixture of amorphous and multiple crystalline phases with different transverse transport properties. Specifically, the amorphous (as-spun) sample has superior $S_{yx}$ and $\Pi_{\text{AEE}}$ values compared to those in the crystalline ($T_{\text{a}} = 1073$ K) sample as shown in Fig. 4f, thus satisfying the conditions for enhanced transverse transport properties in the model. Therefore, the theoretical model also supports our experimentally observed ANE enhancement in the $T_{\text{a}} = 873$ K sample. The red symbols and lines in Figs. 4d-f indicate the samples with large $H_{\text{C}}$, consistent with Figs. 2 and 3.

The summary of this section is as follows. We found an optimized annealing condition for both magnetic and ANE properties (see Fig. S4 in Supplementary Information for the comparison among the sample set); the $T_{\text{a}} = 873$ K sample exhibited large coercivity ($H_{\text{C}} = 750$

Oe), remanence ($M_r/M_s \sim 73\%$), and ANE ($S_{yx} = 2.73 \pm 0.41$ μV/K), while maintaining mechanical flexibility. This suggests the potential application of the sample for energy harvesting from various heat sources with curved or rough surfaces. In the following section, we present the demonstration of an ANE device for energy harvesting from a curved heat source using this material.

## 2.4 Demonstration of ANE-based coiled device applied to curved heat source

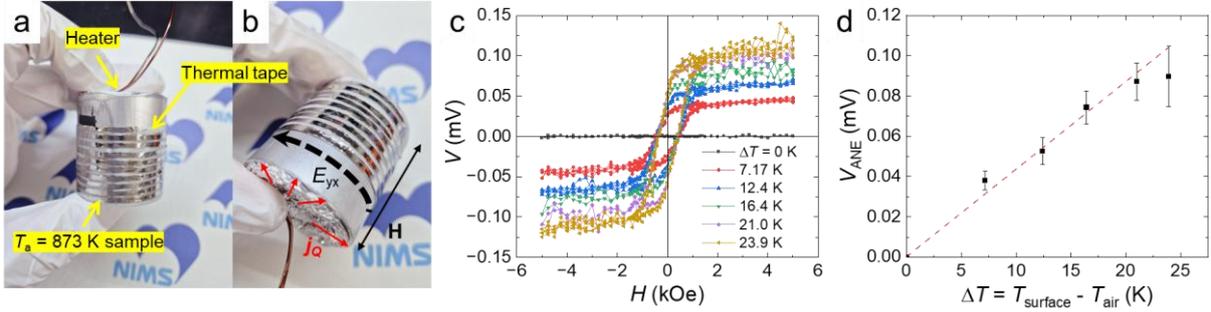

**Fig. 5 | Proof-of-concept demonstration of anomalous-Nernst-effect-based energy harvesting on curved heat source.** (a), (b) Photographs of the coiled device generating a transverse electrical field ($E_{yx}$) under a radial heat flux ($\mathbf{j}_Q$) generated from the pipe-shaped heat source. An external magnetic field (**H**) was applied in the axial direction for quantitative evaluation. (c) Magnetic field dependence of the measured voltage $V$ in the coiled device. The temperature difference $\Delta T$ was estimated by subtracting the ambient air temperature ($T_{air}$ = 295.7 K) from the temperature of the surface ($T_{surface}$). (d) Anomalous Nernst voltage $V_{ANE}$ as a function of $\Delta T$. The dashed line indicates a linear fitting function.

Now we turn our attention to the application of the developed flexible materials with large $H_C$ and $S_{yx}$ for energy harvesting from curved heat sources. We demonstrate this by generating transverse thermoelectric voltage with a coiled device structure using a curved heat source (Figs. 1 and 5a). This structure was previously proposed and demonstrated in ref.[14,36,37] using a single material. These application examples include energy harvesting from heat pipes and automotive applications, both of which require flexible materials or devices. When a heat flux is applied to the radial direction and the material is magnetized in the axial direction, ANE voltage is generated along the tangential direction. This configuration meets the symmetry condition of ANE, and thus enables the cumulative increase in output voltage along with the coil length (Fig. 5b). Therefore, the output voltage and power of the device scale with the material length and area, respectively. This eliminates the needs for additional junctions or n- and p-type material pairs in longitudinal Seebeck-based devices, thus providing a simplified device configuration.

The sample annealed at 873 K was attached to the pipe-shaped heat source with a radius of 19 mm. Thermal tape was used for the attachment, providing electrical insulation between the samples and pipe (Fig. 5a and b). As a result, the coiled material was aligned thermally in parallel and electrically in series. A resistive heater was embedded at the center of a hollow pipe and an aluminum foil was used to dissipate the heat in the radial direction. In addition, an electrical fan was used to decrease the overall temperature of the sample. The surface temperature of the device ($T_{\text{surface}}$) was monitored by a K-type thermocouple attached onto the surface of the pipe (Fig. 5a). The ANE-based transverse thermoelectric conversion was demonstrated under a magnetic field of $\pm 5$ kOe to evaluate the performance of ANE and effective coercivity of the device ($H_{\text{C}}^{\text{eff}}$) quantitatively. Figure 5c shows a clear hysteresis loop of the measured voltage with finite $H_{\text{C}}^{\text{eff}}$ under various values of $\Delta T$ ($= T_{\text{surface}} - T_{\text{air}}$, where $T_{\text{air}}$ indicates the ambient air temperature) at $T_{\text{air}} = 295.7$ K. To clarify the origin of the measured signal, the ANE voltage $V_{\text{ANE}}$ is evaluated using the value at zero magnetic field obtained from the linear extrapolation of data in the saturated field regime (Fig. 5d). The magnitude of $V_{\text{ANE}}$ is linearly proportional to $\Delta T$, indicating that the output voltage originates from the applied radial heat flux, i.e., the transverse thermoelectric conversion due to ANE. The slight decreases in $V_{\text{ANE}}$ at high $\Delta T$ are potentially due to the temperature dependence of $S_{yx}$ of the material because the average temperature of the device increases with increasing $\Delta T$. The $H_{\text{C}}^{\text{eff}}$ value was estimated to be 414 Oe from the voltage measurements in the coiled device, which is smaller than the coercivity of material ($H_{\text{C}}$) observed by VSM (750 Oe). This discrepancy might be attributed to the inhomogeneous formation of crystalline phases in the larger sample used for the device demonstration than those for the VSM and LIT measurements, due to the limited temperature uniformity of the annealing furnace. Nevertheless, we note that this is the largest coercivity reported in flexible ANE materials and devices, typically exhibiting small coercivity of 35-72 Oe[ref.[34,35,40,55,56,58]], and our experimental demonstration proves the applicability of ANE-based flexible devices at zero magnetic field. A detailed comparison is given in the following section, together with $S_{yx}$.

## 2.5 Comparison of magnetic and ANE properties with different materials and devices

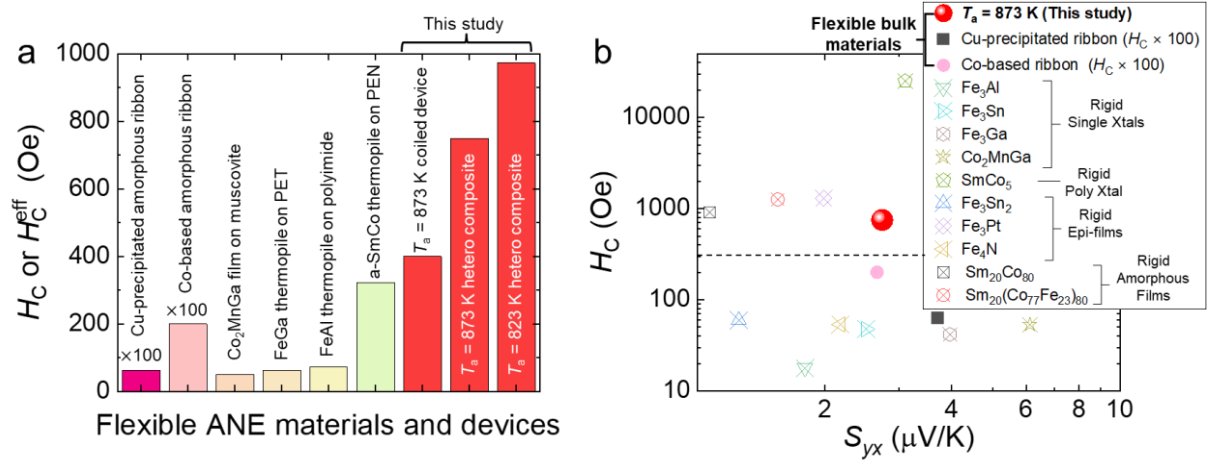

**Fig. 6 | Comparison of coercivity and anomalous Nernst coefficient $S_{yx}$.** (a) Comparison of the coercivity of the coiled device fabricated using the $T_a$ = 873 K sample and the heterogeneous composite samples annealed at $T_a$ = 823 K and 873 K with that of reported flexible magnetic materials and devices, studied for the anomalous Nernst effect. The coercivity values of the Cu-precipitated and Co-based amorphous ribbons are amplified by 100 times for clarity. (b) Comparison of the coercivity for the $T_a$ = 873 K sample with different materials, including both flexible and rigid systems, as a function of $S_{yx}$. The horizontal dashed line in (b) indicates the lower limit of the coercivity used in magnetic stripe cards[39]. The properties of the flexible systems were plotted using the filled symbols in (b). The references for the comprised materials can be found in the main text.

Lastly, we compare the magnetic and transverse thermoelectric properties of the coiled device and the $T_a$ = 873 K and 823 K samples with other systems for quantitative evaluation of their performances (Fig. 6). Figure 6a presents the coercivity values of various flexible magnetic materials ($H_C$) and devices ($H_C^{eff}$). The magnetic thin films and devices fabricated on flexible platforms exhibited small coercivity, ranging from 38-72 Oe[ref.[34,35,55,56]], while flexible amorphous bulk magnetic materials used for ANE studies did small coercivity < 10 Oe[ref.[40,58]]. Notably, our coiled device and bulk samples demonstrated significantly higher coercivity, approximately one or two orders of magnitude larger than the others. These coercivity values exceed that of a magnetic stripe card ($\sim$ 290 Oe [ref.[39]]). The materials comprised in Fig. 6a include an Fe-Ga thermopile on a polyethylene terephthalate (PET) substrate[35], $Co_2MnGa$ film on a muscovite substrate[56], Sm-Co-based amorphous alloy on a polyethylene naphthalate (PEN) substrate[55], Fe-Al thermopile on a polyimide substrate[34], and bulk Cu-precipitated[40] and Co-based amorphous ribbons[58].

Figure 6b shows both $H_C$ and $S_{yx}$ of various materials, encompassing not only flexible materials but also rigid single crystals, polycrystals, and epitaxial thin films. Among these, the $T_a$ = 873 K sample exhibited both large $H_C$ (750 Oe) and $S_{yx}$ (2.73 $\pm$ 0.41 $\mu$V/K) values while maintaining mechanical flexibility, making its performance comparable to that of various rigid systems. The materials included in Fig. 6b are rigid single crystals ($Fe_3Al$[23], $Fe_3Sn$[52], $Fe_3Ga$[23], and $Co_2MnGa$[18–20]), epitaxial thin films on rigid substrates ($Fe_3Sn_2$[53], $Fe_3Pt$[39], and $Fe_4N$[54]), Sm-Co-based amorphous alloys on rigid substrates[57] ($Sm_{20}(Co_{77}Fe_{23})_{80}$ and $Sm_{20}Co_{80}$), and rigid bulk permanent magnet[59] ($SmCo_5$). The comparisons clearly demonstrate the effectiveness of our strategy for designing flexible hard magnets, with their performance surpassing other approaches reported to date.

### 3. Conclusion

We successfully demonstrated the design of flexible hard magnetic materials for zero-magnetic-field operation of ANE, using amorphous-crystalline heterogeneous composites as the platform. Through controlled heat treatments, we optimized the magnetic and transverse thermoelectric properties of the materials, particularly in the $T_a$ = 873 K sample, which exhibited a large coercivity ($H_C$ = 750 Oe) and ANE coefficient ($S_{yx}$ = 2.73 ± 0.41 µV/K). The $H_C$ value is one to two orders of magnitude higher than those reported for other flexible magnetic materials used for ANE studies and the $S_{yx}$ value is comparable to that of rigid magnetic materials showing large ANE. Additionally, the coiled device demonstrated the potential applicability of the flexible hard magnetic materials for energy harvesting from curved heat sources based on ANE. These findings validate our material design strategy and highlight the potential of ANE-based flexible devices in energy harvesting applications, offering a promising alternative to conventional thermoelectric systems.

# 4. Experimental section

## 4.1. Sample preparation

The $Fe_{63}Pt_{15}Si_{12}B_{10}$ master alloy ingot was prepared by melting a mixing of high-purity Fe, B, Pt, and Si elements using a high-frequency induction-heating furnace under an Ar atmosphere. Subsequently, the ingot was crushed into small pieces and charged into a quartz tube with a 5 mm × 0.8 mm nozzle opening. The amorphous ribbons were then produced using the single-roll melt-spinning technique. This process involved melting the ingot pieces using induction heating and ejecting the molten from the nozzle under an Ar atmosphere at a pressure of 0.04 MPa onto a water-cooled Cu wheel rotating at a speed of 30 m/s. The gap between the nozzle and Cu wheel was maintained at 0.2 mm. The amorphous samples were annealed at various temperatures ranging from 673 K to 1073 K under a high vacuum condition ($\sim 10^{-5}$ Pa). Two sample sets were prepared; the main set was annealed at each temperature for 3 min and subsequently quenched into water, while the supplementary set was maintained for 30 min and cooled in the furnace naturally. The heating rate was 20 K/min for both the annealing processes.

## 4.2. Magnetic property analysis

The magnetic properties of the samples were investigated using VSM (Lake Shore, 8600 Series). All the measurements were conducted at room temperature for the magnetic field range of ± 5 kOe.

## 4.3. Structure analysis

The exothermic reaction due to structural change was investigated using DSC (Rigaku, Thermo plus EVO2). The DSC measurement was conducted with a Pt pan at temperatures ranging from 273 K to 993 K with a ramp rate of 10 K/min. The samples were baked at 373 K for 10 min before the measurement to avoid the error from moisture. Material phase analysis was performed using a Rigaku MiniFlex600 X-ray diffractometer with Cr-Kα radiation. Microstructural analysis was conducted using a FEI Titan G2 80-200 TEM. Elemental distribution was observed using APT in a laser mode with a CAMCECA LEAP 5000 XS instrument, operated at a base temperature of 30 K, laser pulse rate of 250 kHz, and energy of 30 pJ. TEM and APT specimens were prepared using the lift-out technique in a FEI Helios 5UX dual beam-focused ion beam. APT data analysis was performed using CAMECA AP Suit 6.1 software.

### 4.4. Characterization of transport properties using lock-in thermography

AEE of all samples was measured using the LIT system (DCG Systems Inc., ELITE). The samples were fixed on a glass substrate and connected electrically in series using silver epoxy. The measurements were conducted at $f$ of 0.5-10 Hz and square-wave-modulated charge current with an amplitude of 100 mA under a magnetic field of $\pm$6.5 kOe at room temperature ($T$ = 300 K).

The thermal diffusivity $D$ was assessed also using the LIT method[40]. The setup includes an infrared camera, diode laser, function generator, and lock-in analysis system. The diode laser, modulated by a reference signal at $f$ from the function generator, directs a focused spot laser beam onto the backside of the sample. This laser spot generates heat waves that spread radially within the sample. As the heat waves propagate, the IR camera captures the thermal response on the front side of the sample. The LIT system processes these thermal images to provide the spatial distribution of the lock-in amplitude $A$ and phase $\phi$ synchronized with $f$. By examining the relationship between $\phi$ and the distance $r$ from the lase point heat source, $D$ can be determined. The formula used for calculating $D$ is

$$D = \frac{\pi f}{\left(\frac{d\phi}{dr}\right)^2}.$$

For the experiment, a high precision LIT system (DCG Systems Inc., ELITE) was used along with a diode laser (Omicron Inc., LDM637D.300.500) with a wavelength of $638 \pm 1$ nm. The laser beam was focused using an optical setup, resulting in an approximately 7-μm spot in a diameter, operating at a power of 20 mW. The beam was modulated at $f = 7.5$ Hz, which was selected to optimize the signal-to-noise ratio and minimize heat losses, ensuring that the thermal diffusion length ($\Lambda = D/\pi f$) remained away from the edges of the sample[69].

### 4.5. Demonstration of ANE-based energy harvesting using coiled structure

The sample annealed at 873 K for 3 min was used for the device demonstration owing to its large $H_C$ and $S_{yx}$ with mechanical flexibility. The sample was attached onto the hollow pipe with a radius of ~19 mm using a thermally conductive and electrically insulating adhesive tape (3M, VHR0601-03). The total length of the device was approximately 1 m. A 40 Ω heater was embedded at the center of the pipe and an aluminum foil was used to dissipate the heat from the heater to the pipe in the radial direction. A DC power supply (TEXIO, PA36-3B) applied a charge current to the heater to generate the radial heat flux. The surface temperature of the device was monitored by a K-type thermocouple attached onto the surface of the pipe using a thermal tape with a digital multimeter (Keithley, 2000). The external magnetic field was applied using a customized superconducting magnet (Niki Glass Company Ltd., Model LTS-408-RTB-100) for the magnetic field range of $\pm$ 5 kOe. The non-uniformity of the magnetic field along the device was less than 5 %, confirmed using a Hall sensor. The ANE voltage was measured at room temperature ($T_{air}$ = 295.7 K) using a nanovoltmeter (Keithley, 2182a). The effective coercivity of the device ($H_C^{eff}$) was estimated from Fig. 5c, corresponding to the magnetic field at which the ANE voltage reaches zero.

## Author Contributions

S.J.P. designed and conceived the study. K.U. supervised the project. H.S.A. fabricated the initial as-spun amorphous metal. S.J.P. conducted the heat treatment of the samples with help of R.G., measured and analyzed the DSC and VSM data with help of R.M., performed the LIT measurements and analyzed the data with help of T.H, F.A. and K.U., and fabricated the coiled device and performed the ANE measurement with help of R.M. R.G. and H.S.A. performed the XRD, TEM, and APT experiments and investigated the structural transition in the samples. A.A. and H.N. measured the thermal diffusivity of the samples using LIT. S.J.P. wrote the manuscript with input from all authors.

## Conflict of Interest

The authors declare no conflict of interest.

## Data availability statement

The data that support the findings of this study are available from the corresponding author upon reasonable request.

## Acknowledgements

The authors thank Ki Mun Bang, Jangwoo Ha, and Hyungyu Jin at POSTECH for their fruitful discussions during the early stages of idea development and Mizue Isomura and Sebata Hiroyuki for technical supports. This work was supported by ERATO "Magnetic Thermal

Management Materials" (grant no. JPMJER2201) from JST, Japan.

# Supplementary Information of

# Designing flexible hard magnetic materials for zero-magnetic-field operation of the anomalous Nernst effect

Sang J. Park[1,*], Rajkumar Modak[1,#], Ravi Gautam[1], Abdulkareem Alasli[2], Takamasa Hirai[1], Fuyuki Ando[1], Hosei Nagano[2], Hossein Sepehri-Amin[1] and Ken-ichi Uchida[1,3,*]

[1] National Institute for Materials Science, Tsukuba 305-0047, Japan

[2] Department of Mechanical Systems Engineering, Nagoya University, Nagoya 464-8601, Japan

[3] Department of Advanced Materials Science, Graduate School of Frontier Sciences, The University of Tokyo, Kashiwa 277-8561, Japan

*Correspondence to: PARK.SangJun@nims.go.jp (S.J.P.);

UCHIDA.Kenichi@nims.go.jp (K.U.)

[#]Present affiliation: Department of Advanced Materials Science, Graduate School of Frontier Sciences, The University of Tokyo, Kashiwa 277-8561, Japan

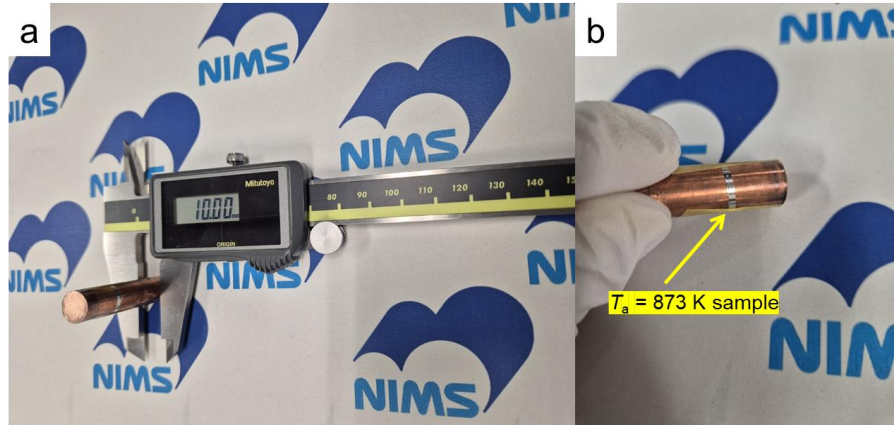

**Fig. S1 |** Photographs of a curved surface with a radius of 5 mm (a) and the $T_a$ = 873 K sample, optimized for both coercivity and anomalous Nernst coefficient, attached to the curved surface (b). A Cu rod was used for the flexibility test. $T_a$ denotes the annealing t emperature.

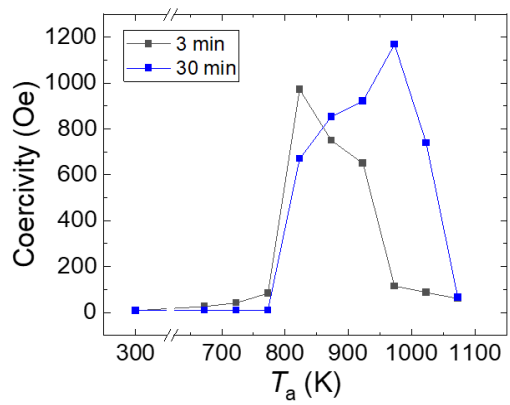

**Fig. S2 |** Coercivity ($H_C$) of the samples, including both 3 min and 30 min annealed samples, as a function of the annealing temperature.

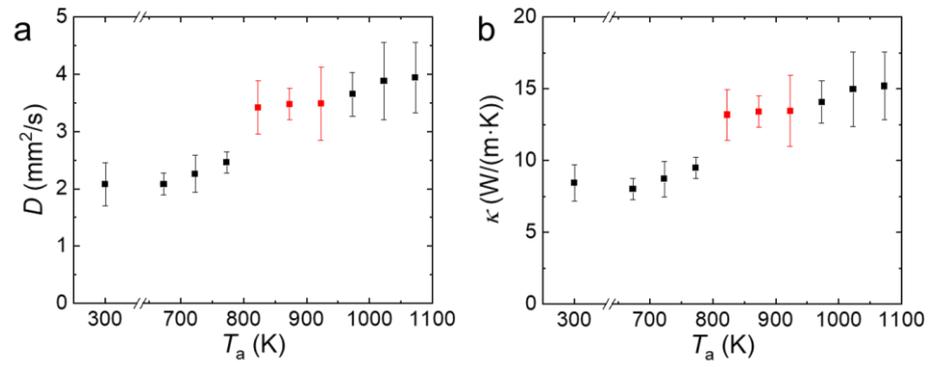

**Fig. S3 |** Measured thermal diffusivity $D$ (a) and thermal conductivity $\kappa$ estimated from $D$ (b) as a function of $T_a$. The red symbols in (a) and (b) correspond to the samples exhibiting a significant enhancement of the coercivity.

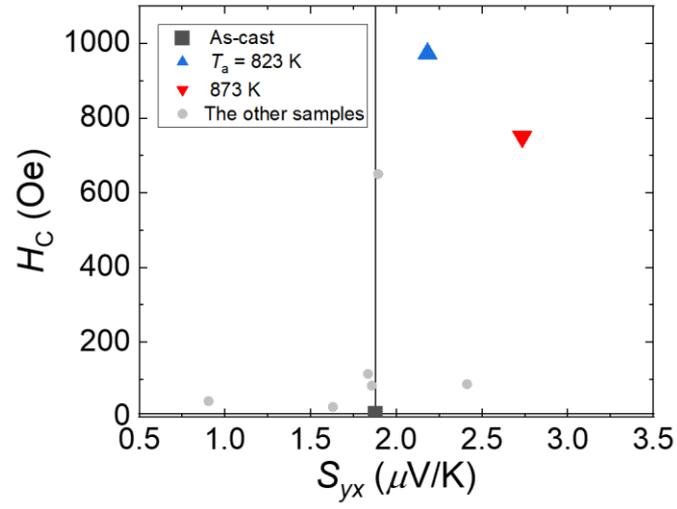

**Fig. S4 |** Coercivity vs. anomalous Nernst coefficient ($S_{yx}$) in the samples annealed under various temperatures. The $T_a$ = 823 and 873 K samples show a superior performance compared to the other samples, annealed at 673, 723, 773, 923, 973, 1023, and 1073 K.